\input{aipcheck}

\documentclass[
    ,draft            
  ]
  {aipproc}

\layoutstyle{6x9}

\usepackage{amsmath, amsfonts,amssymb,amsthm, bbm}
\usepackage{braket}

\usepackage{graphicx}   
\usepackage{subfigure}  

\newcommand\abs[1]{|#1|}
\newcommand\argmin[1]{\underset{#1}{\operatorname{argmin}}}

\begin{document}

\title[ADAPTIVE HAMILTONIAN ESTIMATION]{ADAPTIVE HAMILTONIAN ESTIMATION USING BAYESIAN EXPERIMENTAL DESIGN}


\author{Christopher Ferrie}{
  address={Institute for Quantum Computing, University of Waterloo, Waterloo, Ontario, Canada},
  altaddress={Department of Applied Mathematics, University of Waterloo, Ontario, Canada},
  email={csferrie@uwaterloo.ca}
}

\author{Christopher E. Granade}{
  address={Institute for Quantum Computing, University of Waterloo, Waterloo, Ontario, Canada},
  altaddress={Department of Physics, University of Waterloo, Ontario, Canada},
  email={cgranade@cgranade.com}
}

\author{D. G. Cory$^{\ddag,}$}{
  address={Institute for Quantum Computing, University of Waterloo, Waterloo, Ontario, Canada},
  altaddress={Perimeter Institute for Theoretical Physics, Waterloo, Ontario, Canada},
  altaddress={Department of Chemistry, University of Waterloo, Ontario, Canada}
}

\begin{abstract}
Using Bayesian experimental design techniques, we have shown that for a single two-level quantum mechanical system under strong (projective) measurement, the dynamical parameters of a model Hamiltonian can be estimated with exponentially improved accuracy over offline estimation strategies.  To achieve this, we derive an adaptive protocol which finds the optimal experiments based on previous observations.  We show that the risk associated with this algorithm is close to the global optimum, given a uniform prior.  Additionally, we show that sampling at the Nyquist rate is not optimal.
\end{abstract}

\maketitle

\section{Introduction\label{section:intro}}
Quantum mechanics gives the most accurate description of many physical systems of interest.  In turn, the most accurate characterization of a quantum device is given by its quantum mechanical model.  Thus, efficient methods for the honest estimation of the distribution of parameters in a quantum mechanical model are of utmost importance, not only for building robust quantum technologies, but to reach new regimes of physics.

Bayesian experimental design (see, e.g. \cite{Loredo2004Bayesian}) is a methodology to ascertain the utility of a proposed experiment.
Bayesian experimental design has been successfully applied to problems in experimental physics, such as in the recent examples of \cite{dreier_bayesian_2008} and \cite{von_toussaint_optimizing_2010}.
In classical theories of physics and statistics, the measurement simply reveals the state of the system at that instant.  By contrast, quantum theory presents with the following physical (and conceptual) barrier: no single measurement can reveal the state.  Rather, each potential kind of experiment admits a probability distribution from which we draw our data. Thus, the methodology of experimental design seems tailor-made for quantum theory.

The structure of the paper is as follows.  We begin by reviewing the general outline of Bayesian experimental design.  We then apply the technique to devise an algorithm for the estimation of quantum Hamiltonian parameters.  We show that in a particular case, this strategy is nearly globally optimal and demonstrate its improvement over standard algorithms numerically.  Finally we conclude with a discussion on the applicability of this technique to real experiments on more complex quantum systems.

\section{Bayesian experimental design\label{section:BED}}

We assume some initial experiment $E$ has been performed and data $D$ has been obtained.  The goal is to determine $\Pr(\Theta|D,E)$, the probability distribution of the model parameters $\Theta$ given the experimental data.  To achieve this we use Bayes' rule
\[
\Pr(\Theta|D,E)=\frac{\Pr(D|\Theta,E)\Pr(\Theta|E)}{\Pr(D|E)},
\]
where $\Pr(D|\Theta,E)$ is the \emph{likelihood function}, which is determined through the process of modeling the experiment, and $\Pr(\Theta|E)$ is the \emph{prior}, which encodes any \emph{a priori} knowledge of the model parameters.  The final term $\Pr(D|E)$ can simply be thought as a normalization factor.

At this stage we can stop or obtain further data.  Experimental design is well suited to quantum theory since an arbitrary fixed measurement procedure does not give maximal knowledge as is often assumed in the statistical modeling of classical system.  We conceive, then, of possible future data $D_1$ obtained from a, possibly different, experiment $E_1$.  The probability of obtaining this data can be computed from the distributions at hand via marginalizing over model parameters
\[
\Pr(D_1|E_1, D, E)=\int  \Pr(D_1|\Theta,E_1)\Pr(\Theta|D,E) d\Theta.
\]
We can use this distribution to calculate the expected \emph{utility} of an experiment
\[
U(E_1)=\sum_{D_1} \Pr(D_1|E_1, D, E) U(D_1,E_1),
\]
where $U(D_1,E_1)$ is the utility we would derive if experiment $E_1$ gave result $D_1$.  This could in principle be any function tailored to the specific problem.  However, for scientific inference, a generally well motivated measure of utility is \emph{information gain} \cite{Lindley1956Measure}.  In information theory, information is measured by the entropy
\[
U(D_1,E_1)=\int \Pr(\Theta|D_1,E_1,D, E) \log\Pr(\Theta|D_1,E_1,D, E) d\Theta.
\]
Thus, we search for the experiment which maximizes the expected information in the final distribution.  That is, an optimal experiment $\hat E$ is one which satisfies
\begin{align*}
&U(\hat E) =\max_{E_1} \Big\{\sum_{D_1} \Pr(D_1|E_1, D, E) \times \\
&\int \Pr(\Theta|D_1,E_1,D, E) \log\Pr(\Theta|D_1,E_1,D, E) d\Theta\Big\}.
\end{align*}

\section{Application to Simple Example\label{section:simpleEx}}

As an example of how to apply the Bayesian experimental design formalism to problems in quantum information,
we consider a simple situation with a single qubit. In particular, we suppose that the qubit evolves under an
internal Hamiltonian
\[
H = \frac{\omega}2 \sigma_z.
\]
Here $\omega$ is an unknown parameter whose value we want to estimate.  An experiment consists of preparing a single known input state $\psi_{\text{in}}=\ket{+}$, the $+1$ eigenstate of $\sigma_x$, evolving under the Hamiltonian $H$ for a controllable time $t$ and performing a measurement in the $\sigma_x$ basis.  This is the simplest problem where adaptive Hamiltonian estimation can be used and is the problem studied in reference \cite{Sergeevich2011Characterization}.

In the language of Bayesian inference, the data $D\in\{0,1\}$ is the outcome of the measurement.  An experiment $E$ consists of a specification of time the $t$ that the Hamiltonian is on, while the model parameter $\Theta$ is simply $\omega$.  The likelihood function is given by the Born rule
\[
\Pr(D=0|\Theta,E) = \abs{\bra{+}e^{i\frac{\omega}2\sigma_zt}\ket{+}}^2 = \cos^2(\frac\omega2 t).
\]
Experimental design is a decision theoretic problem based on the utility function
\[
U(t)=\sum_{D} \Pr(D|t) \int \Pr(\omega|D,t) \log\Pr(\omega|D,t) d\omega.
\]
The optimal design is any value of $t$ which maximizes this quantity.

\begin{figure}
  \includegraphics[width=.9\linewidth]{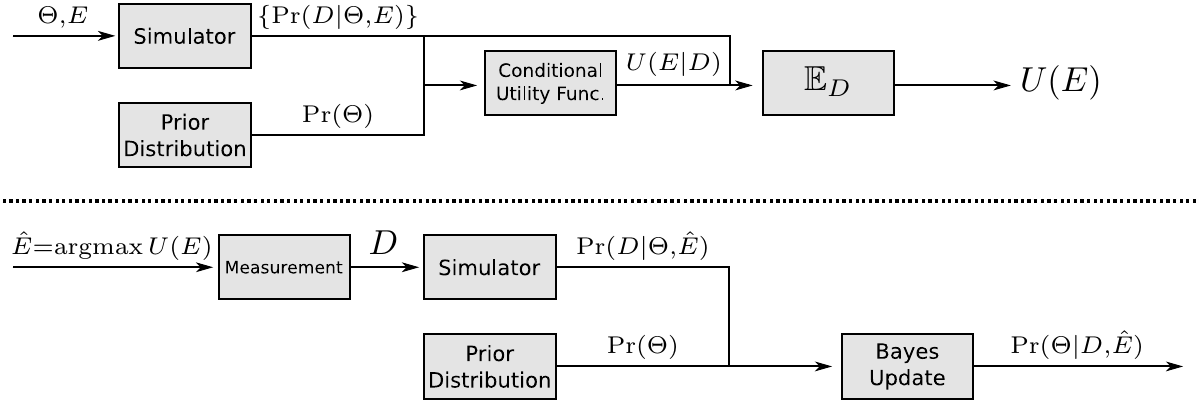}
  \caption{\label{fig:overview} Overview of a step in the online adaptive algorithm for finding locally optimal experiments. Top: Method for calculating the utility function $U(E)$, given a simulator and a prior distribution $\Pr(\Theta)$ over model parameters $\Theta$. Bottom: Method for updating prior distribution with results $D$ from chosen actual experiment.}
\end{figure}

We proceed by performing the optimal experiment and obtaining data $D_1$.  Using Bayesian inference we update our prior $\Pr(\omega)$ via Bayes' rule:
\[
\Pr(\omega|D_1)=\frac{\Pr(D_1|\omega)\Pr(\omega)}{\Pr(D_1)}.
\]
If we are not satisfied, we can repeat the process where this distribution becomes the prior for the new experimental design step.  This algorithm is depicted in figure \ref{fig:overview}.

\subsection{Estimators, squared error loss and a greedy alternative to information gain\label{section:MSE}}

The preceding problem had a single unknown variable.  If we desire an estimate $\hat\Theta$ of the true value $\Theta$, the most often used figure of merit is the \emph{squared error loss}:
\[
L(\Theta,\hat \Theta) = \abs{\Theta-\hat\Theta}^2.
\]

The \emph{risk} of an estimator $\hat\Theta:\{D,D_1,D_2,\ldots,D_N\}\mapsto \mathbb R$ is its expected performance with respect to the loss function:
\[
R(\Theta,\hat\Theta) = \sum_{\{D,D_1,D_2,\ldots,D_N\}} \Pr(\{D,D_1,D_2,\ldots,D_N\}|\Theta) L(\Theta,\hat\Theta).
\]
For squared error loss, the risk is also called the \emph{mean squared error}.  The average of this quantity with respect to some prior $\Pr(\Theta)=:\pi(\Theta)$ is the \emph{Bayes risk} of $\pi$,
\[
r(\pi,\hat\Theta) = \int R(\Theta,\hat\Theta) \pi(\Theta) d\Theta,
\]
and the estimator which minimizes this quantity is called a \emph{Bayes estimator}.  In this case the Bayes estimator is the mean of the posterior distribution\footnote{Note that in any case where the loss function is \emph{strictly proper}, i.e. is equal to zero if and only if the estimate is equal to the true state, the Bayes estimator is the posterior mean \cite{BlumeKohout2006Accurate}.}.  Let us assume then that the estimators we choose are Bayes.  Let us also choose a uniform prior for $\Theta$.  Then, the final figure of merit is the average mean squared error (AMSE):
\[
r = \int R(\Theta,\hat\Theta) d\Theta.
\]
We would like a strategy which minimizes this quantity.  Non-adaptive Fourier and Bayesian strategies were investigated and compared to an adaptive strategy in reference \cite{Sergeevich2011Characterization}.  Their adaptive strategy fits into the Bayesian experimental design framework when the utility is measured by the \emph{variance} of the posterior distribution:
\[
V(D_1,E_1) = - \int \Pr(\Theta|D_1,E_1,D, E)  (\Theta^2 - \mu(D_1,E_1))^2 d\Theta,
\]
where
\[
\mu(D_1,E_1) = \int \Pr(\Theta|D_1,E_1,D, E) \Theta d\Theta
\]
is the mean of the posterior.  Recall that the mean is a Bayes estimator of AMSE, so $\mu=\hat\Theta$.  For a single measurement this utility function satisfies $V = -r$.  That is, maximizing the utility \emph{locally} at each step of the algorithm is equivalent to minimizing the AMSE at each step. Hence, when using the negative variance as our utility function, the adaptive strategy summarized in Figure \ref{fig:overview} is an example not only of a local optimization, but also a \emph{greedy algorithm} with respect to the AMSE risk. 
In the future, we shall refer to this choice of utility function together with the local optimization algorithm as the greedy algorithm for this problem.

We can write the risk of this strategy recursively as follows.  Suppose at the $N$'th, and final, measurement we have the updated distribution $\pi_{N-1}$.  Then, the risk of the local strategy is
\[
l_N(\pi_{N-1},\Theta) =  \sum_{D_N} \Pr(D_N|\Theta,\hat E_N) L(\Theta,\mu(D_N,\hat E_N)),
\]
where $\hat E_N$ is the locally optimal design satisfying
\[
\hat E_N = \argmin{E_N} \int \sum_{D_N} \Pr(D_N|\Theta,E_N) L(\Theta,\mu(D_N,E_N))) \pi_{N-1}(\Theta) d\Theta.
\]
The expected risk at any other stage is
\[
l_n(\pi_{n-1},\Theta) = \sum_{D_n} \Pr(D_n|\hat E_n) l_{n+1}\left(\frac{\Pr(D_n|\Theta,\hat E_n)\pi_{n-1}(\Theta)}{\int\Pr(D_n|\Theta,\hat E_n)\pi_{n-1}(\Theta)d\Theta}\right),
\]
where $\hat E_n$ is, again, the locally optimal design satisfying
\[
\hat E_n = \argmin{E_n} \int \sum_{D_n} \Pr(D_n|\Theta,E_n) L(\Theta,\mu(D_n,E_n))) \pi_{n-1}(\Theta) d\Theta.
\]
Then, the Bayes risk of the greedy strategy is
\[
\int l_1(\pi_0,\Theta) \pi_0(\Theta)d\Theta.
\]

Again, it is clear that the greedy algorithm is globally optimal on the final decision, as there is no further hypothetical data to consider.  That is, the optimal solution at the $N$'th measurement is
\[
g_N(\pi_{N-1},\Theta) =  \sum_{D_N} \Pr(D_N|\Theta,\hat E_N) L(\Theta,\mu(D_N,\hat E_N)),
\]
where $\hat E_N$ is the locally optimal design satisfying
\[
\hat E_N = \argmin{E_N} \int \sum_{D_N} \Pr(D_N|\Theta,E_N) L(\Theta,\mu(D_N,E_N))) \pi_{N-1}(\Theta) d\Theta.
\]
However, the globally optimal risk at any other stage
\[
g_n(\pi_{n-1},\Theta) = \sum_{D_n} \Pr(D_n|\tilde E_n) g_{n+1}\left(\frac{\Pr(D_n|\Theta,\tilde E_n)\pi_{n-1}(\Theta)}{\int\Pr(D_n|\Theta,\tilde E_n)\pi_{n-1}(\Theta)d\Theta}\right),
\]
where now $\tilde E_n$ is the globally optimal design satisfying
\[
\tilde E_n = \argmin{E_n} \int \sum_{D_n} \Pr(D_n|\Theta,E_n) g_{n+1}\left(\frac{\Pr(D_n|\Theta, E_n)\pi_{n-1}(\Theta)}{\int\Pr(D_n|\Theta,E_n)\pi_{n-1}(\Theta)d\Theta}\right)\pi_{n-1}(\Theta) d\Theta.
\]
Then, the Bayes risk of the greedy strategy is
\[
\int g_1(\pi_0,\Theta) \pi_0(\Theta)d\Theta.
\]

In general, $l_1(\pi_0,\Theta) \neq g_1(\pi_0,\Theta)$.  Nor is it the case that
\[
\int l_1(\pi_0,\Theta) \pi_0(\Theta)d\Theta = \int g_1(\pi_0,\Theta) \pi_0(\Theta)d\Theta
\]
for an arbitrary prior.  However, for the special case of the uniform prior, we have found numerically that the Bayes risk of the greedy strategy and the Bayes risk of the global strategy are similar enough that the greedy strategy is useful.

\subsection{Performance comparisons\label{section:simple results}}

In reference \cite{Sergeevich2011Characterization}, it was shown via simulation that the posterior variance of the greedy strategy is best fit by an exponentially decreasing function of $N$, the total number of measurements.  In contrast, all off-line strategies decrease at best as a linear function of $N$.

\begin{figure}
  \includegraphics[width=.49\linewidth]{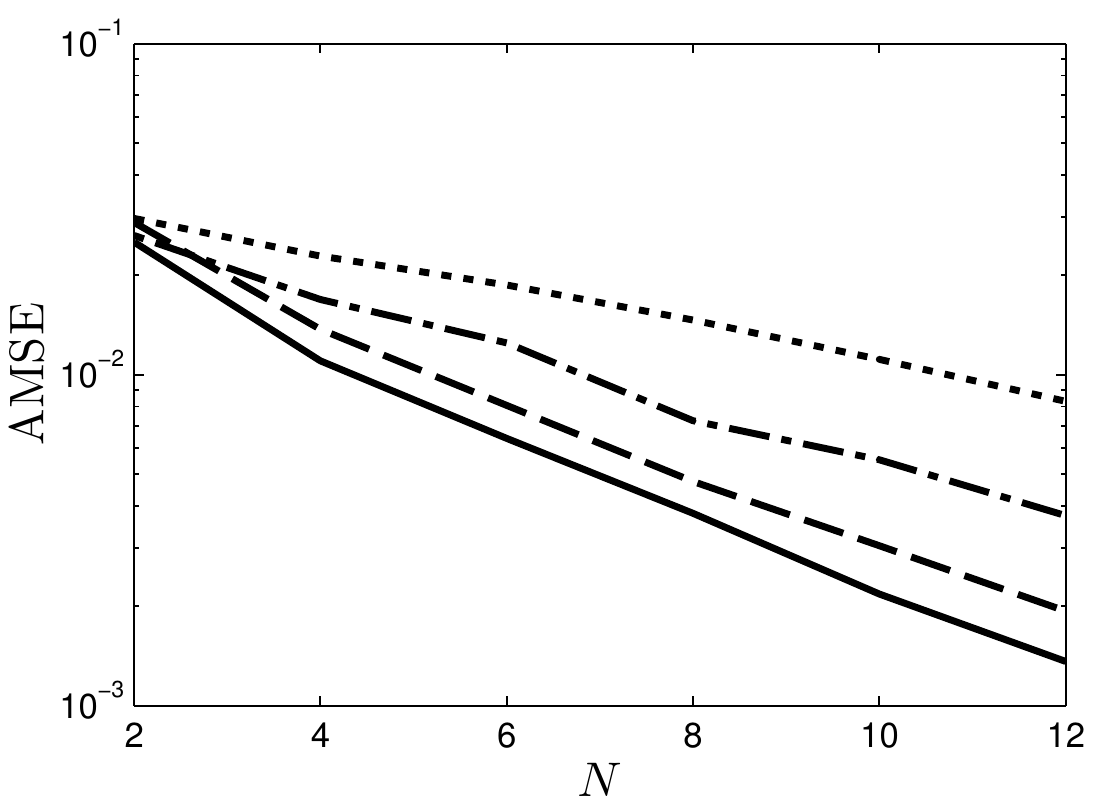}  \includegraphics[width=.49\linewidth]{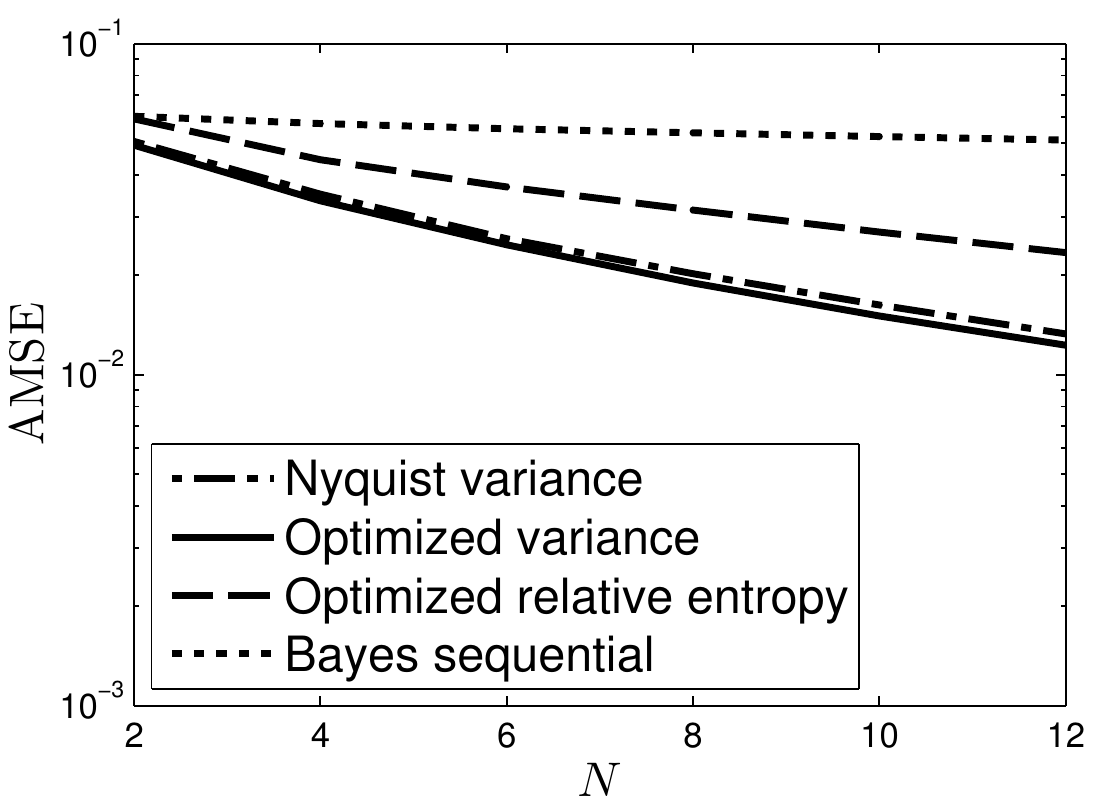}
  \caption{\label{fig:results} Performance of the estimation strategies.  The Bayesian sequential and the strategy labeled ``Nyquist'' sample at the Nyquist rate.  The ``optimized'' strategies find the global maximum utility (using Matlab's ``fmincon'' starting with the optimal Nyquist time). In each case, $N_{\max} = 12$ measurements are considered.  Left: the ideal model discussed in the text.  Right: a more realistic model with $25\%$ noise and an addition relaxation process (known as $T_2$) which exponentially decays the signal (to half its value at $t=10\pi$). }
\end{figure}

In Figure \ref{fig:results}, we show that the local information gain optimizing algorithm also enjoys an exponential improvement in accuracy over naive off-line methods.  Moreover, we show Nyquist rate sampling is unnecessary and, indeed, sub-optimal.  All results stated are obtained using a uniform prior on $[0,1]$ and are computed numerically by exploring every branch of the the decision tree, in contrast to simulation.

In order to be ``fair'' to the off-line methods, we restricted the adaptive methods to explore the same experimental design specifications.  That is, for this particular problem, the adaptive algorithm was allowed to select measurement times from $[0,N_{\text{max}}\pi]$, where $N_{\text{max}}$ is the total number of measurements.  In principle, these methods could only do better with a larger design specification.

\section{Discussion\label{section:conclusions}}

Summarizing, we have shown for the problem of estimating the parameter in a simple Hamiltonian model of qubit dynamics an adaptive measurement strategy can exponentially improve the accuracy over offline estimation strategies.  Moreover, we have shown that sampling at the Nyquist rate is not optimal in the case of strong measurement.  We have derived a recursive solution to the risks for both the local and global optimal strategies. Using this solution, we numerically found that the local strategy is nearly optimal in the special case of a uniform prior. That the greedy algorithm is nearly optimal in a case relevant to experiment demonstrates that an adaptive Bayesian method may be computationally feasible, in that an implementation need not consider all possible future data when choosing each experiment.

Together, these results demonstrate the usefulness of an adaptive
Bayesian algorithm for parameter estimation in quantum
mechanical systems, especially in comparison with other
algorithms in common use. In the presence of noise, this
improvement becomes still more stark, as demonstrated by the results shown in Figure \ref{fig:results}.

\begin{figure}
  \includegraphics[width=.49\linewidth]{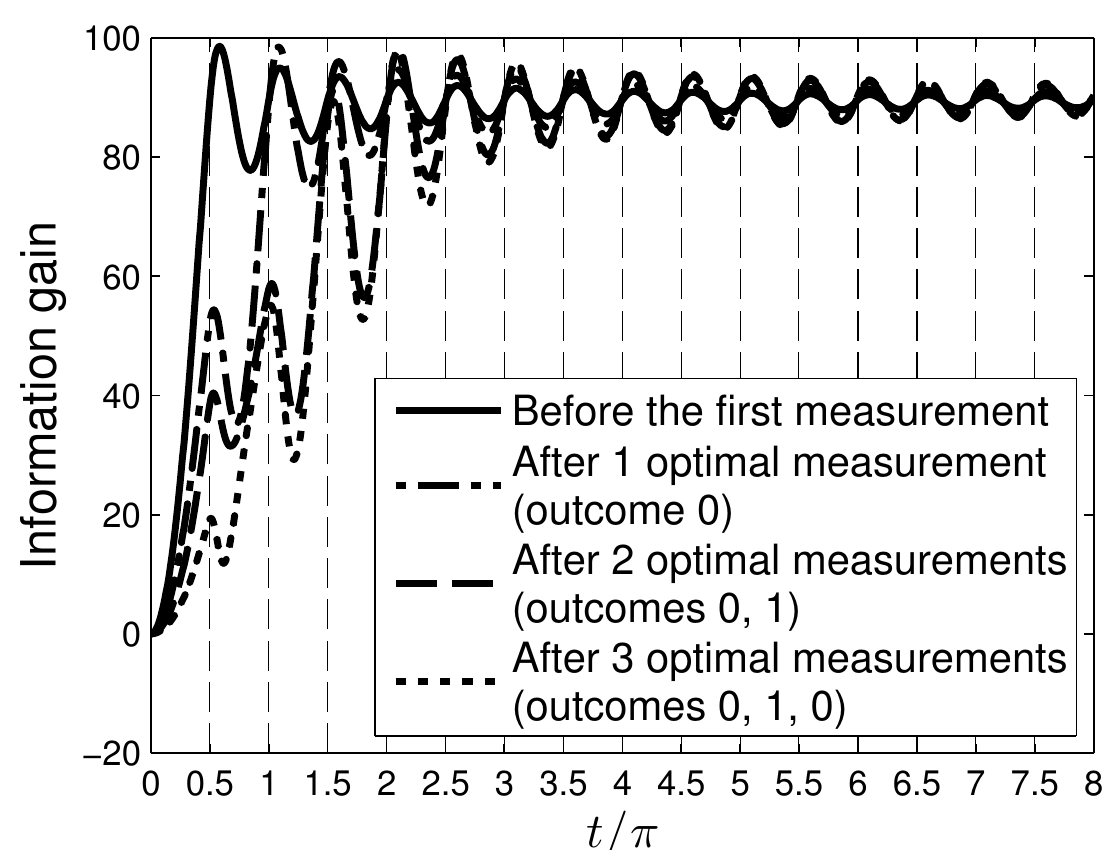}
  \includegraphics[width=.49\linewidth]{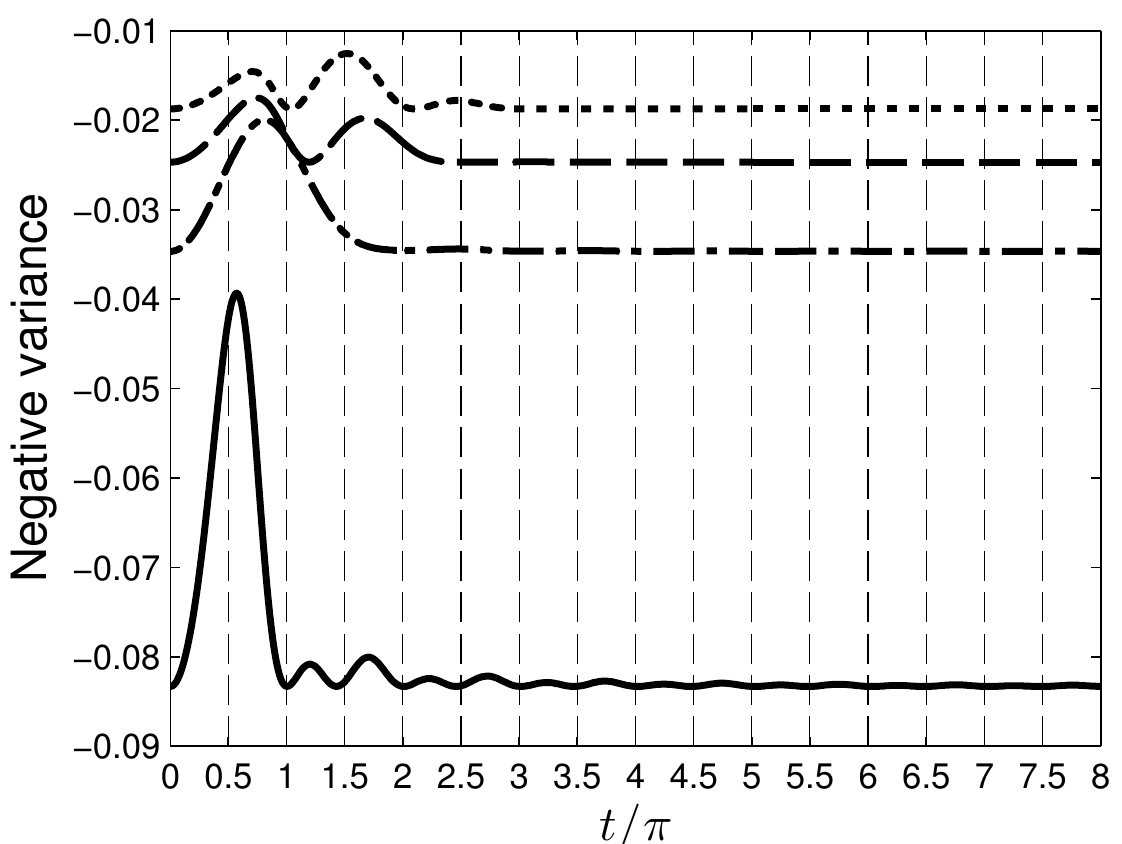}
  \caption{\label{fig:utils} The information gain (left) and variance (right) utilities for the prior followed by three simulated measurements.  The vertical grid lines indicate the Nyquist times. Note that the times at which the utilities are maximized do not necessarily increase with the number of measurements.}
\end{figure}

Why is it the case that the Nyquist times are not optimal?  First, why should we expect them to be optimal?  The Nyquist theorem states that a signal which contains no frequencies higher than $\omega_{\text{max}}$ is completely and unambiguously characterized by a discrete set of samples taken at a rate greater than or equal to $\pi/2\omega_{\text{max}}$.  However, the classical notion of sampling fails for the strong-measurement case that we consider here.  What we have is a periodic \emph{probability} distribution which can be sampled, not a periodic function whose \emph{values} can be ascertained.  That is there is no \emph{signal}, in the classical sense of the word, which can be reconstructed.  The failure of the Nyquist rate sampling is exemplified in Figure \ref{fig:utils}.

In this paper, we have chosen to measure success via the squared error loss.  Although this is a standard metric, note that it is not practically useful in the context of estimating the parameters of a quantum mechanical system.  We motivate this claim as follows.  A typical application of our algorithm is to inform control theory algorithms, which can achieve significantly higher fidelities if given a distribution over Hamiltonians rather than a single best estimate.
Indeed, in the case of nuclear magnetic resonance, the physical ensemble of qubits produces a real distribution of Hamiltonians to which control theory algorithm must be robust against \cite{Boulant2003Robust,Boulant2004Incoherent}. Any single estimate of the Hamiltonian parameters will thus artificially exclude dynamics which will appear as decoherence in the resultant pulses. Thus, we must measure the success of our algorithm via a loss function of the true \emph{distribution} and estimated posterior.  Noting that relative entropy is broadly considered the correct loss function for probability estimators, our algorithm, which maximizes expected information gain, becomes the optimal solution.

We expect that in more complicated systems, the Bayesian adaptive
method will remain useful, especially in applications such as
optimal control theory, where having a distribution over
Hamiltonians is significantly more useful than a single best
estimate.




\begin{theacknowledgments}
  CF thanks Josh Combes for helpful discussions.  This work was financially supported by NSERC and CERC.
\end{theacknowledgments}

\end{document}